\documentclass[runningheads]{llncs}
\usepackage{graphicx}
\usepackage{amssymb}
\usepackage{amsmath,tabularx}
\usepackage{xcolor}
\usepackage{multirow}
\usepackage{siunitx}
\usepackage{booktabs}
\usepackage{etoolbox}

\begin{document}

\robustify\bfseries
        
\title{TeCNO: Surgical Phase Recognition with Multi-Stage Temporal Convolutional Networks}

\titlerunning{Surgical Phase Recognition with Temporal Convolutional NNs}

\author{Tobias Czempiel\inst{1}, Magdalini Paschali\inst{1}, Matthias Keicher\inst{1}, Walter Simson\inst{1}, Hubertus Feussner\inst{2}, Seong Tae Kim\inst{1}, Nassir Navab\inst{1,3}}

\authorrunning{Czempiel et al.}

\institute{
Computer Aided Medical Procedures, Technische Universit{\"a}t M{\"u}nchen, Germany
\and
MITI, Klinikum Rechts der Isar, Technische Universit{\"a}t M{\"u}nchen, Germany
\and
Computer Aided Medical Procedures, Johns Hopkins University, Baltimore, USA 
}

\maketitle              
\begin{abstract}
Automatic surgical phase recognition is a challenging and crucial task with the potential to improve patient safety and become an integral part of intra-operative decision-support systems. In this paper, we propose, for the first time in workflow analysis, a Multi-Stage Temporal Convolutional Network (MS-TCN) that performs hierarchical prediction refinement for surgical phase recognition. Causal, dilated convolutions allow for a large receptive field and online inference with smooth predictions even during ambiguous transitions. Our method is thoroughly evaluated on two datasets of laparoscopic cholecystectomy videos with and without the use of additional surgical tool information. Outperforming 
various state-of-the-art LSTM approaches, we verify the suitability of the proposed causal MS-TCN for surgical phase recognition.

\end{abstract}

\section{Introduction} 
Surgical workflow analysis is an integral task to increase patient safety, reduce surgical errors and optimize the communication in the operating room (OR)~\cite{or_next_generation}. Specifically, surgical phase recognition can provide vital input to physicians in the form of early warnings in cases of deviations and anomalies~\cite{Huaulme2020} as well as context-aware decision support~\cite{Padoy2019}. Another use case is automatic extraction of a surgery's protocol, which is crucial for archiving, educational and post-operative patient-monitoring purposes~\cite{Zisimopoulos2018}.

Computer-assisted intervention (CAI) systems based on machine learning techniques have been developed for surgical workflow analysis~\cite{Padoy2012}, deploying not only OR signals but also intra-operative videos, which can be captured during a laparoscopic procedure, since cameras are an integral part of the workflow. However, the task of surgical phase recognition from intra-operative videos remains challenging even for advanced CAI systems~\cite{Lecuyer2020},~\cite{Bodenstedt2019} due to the variability of patient anatomy and surgeon style~\cite{Funke2019} along with the limited availability and quality of video material~\cite{Klank2008}. Furthermore, strong similarities among phases and transition ambiguity lead to decreased performance and limited generalizability of the existing methods. Finally, most approaches dealing with temporal information, such as Recurrent Neural Networks (RNNs)~\cite{AlHajj2019} leverage sliding window detectors, which have difficulties capturing long-term temporal patterns.

Towards this end, we propose a pipeline utilizing dilated Temporal Convolutional Networks (TCN)~\cite{Lea} for accurate and fast surgical phase recognition. Their large temporal receptive field captures the full temporal resolution with a reduced number of parameters, allowing for faster training and inference time and leveraging of long, untrimmed surgical videos.

Initial approaches for surgical phase recognition~\cite{Padoy2012} exploited binary surgical signals. Hidden Markov Models (HMMs) captured the temporal information with the use of Dynamic Time Warping (DTW). However, such methods relied on whole video sequences and could not be applied in an online surgery scenario.
EndoNet~\cite{Twinanda2017} jointly performed surgical tool and phase recognition from videos, utilizing a shared feature extractor and a hierarchical HMM to obtain temporally-smoothed phase predictions. 
With the rise of RNNs, EndoNet was evolved to EndoLSTM, which was trained in a two-step process including a Convolutional Neural Network (CNN) as a feature extractor and an LSTM~\cite{doi:10.1162/neco.1997.9.8.1735} for feature refinement. Endo2N2~\cite{Yengera2018} leveraged self-supervised pre-training of the feature extractor CNN by predicting the Remaining Surgery Duration (RSD). Afterwards a CNN-LSTM model was trained end-to-end to perform surgical phase recognition. Similarly, SV-RCNet~\cite{Jin2018} trained an end-to-end ResNet~\cite{He} and LSTM model for surgical phase recognition with a prior knowledge inference scheme.

MTRCNet-CL~\cite{Jin2020} approached surgical phase classification as a multi-task problem.
Extracted frame features were used to predict tool information while also serving as input to an LSTM model~\cite{doi:10.1162/neco.1997.9.8.1735} for the surgical phase prediction. A correlation loss was employed to enhance the synergy between the two tasks.
The common factor of the methods mentioned above is the use of LSTMs, which retain memory of a limited sequence, that cannot span minutes or hours, which is the average duration of a surgery. Thus, they process the temporal information in a slow, sequential way prohibiting inference parallelization, which would be beneficial for their integration in an online OR scenario. 

Temporal convolutions~\cite{Lea} were introduced to hierarchically process videos for action segmentation. An encoder-decoder architecture was able to capture both high- and low-level features in contrast to RNNs. Later, TCNs adapted dilated convolutions~\cite{VanDenOord} for action localization and achieved improvement in performance due to a larger receptive field for higher temporal resolution. Multi-Stage TCNs (MS-TCNs)~\cite{Farha} were introduced for action segmentation and consisted of stacked predictor stages. Each stage included an individual multi-layer TCN, which incrementally refined the initial prediction of the previous stages.

In this paper our contribution is two-fold:
(1) We propose, for the first time in surgical workflow analysis, the introduction of causal, dilated MS-TCNs for accurate, fast and refined online surgical phase recognition. We call our method TeCNO, derived from \textbf{Te}mporal \textbf{C}onvolutional \textbf{N}etworks for the \textbf{O}perating room.
(2) We extensively evaluate TeCNO on the challenging task of surgical phase recognition on two laparoscopic video datasets, verifying the effectiveness of the proposed approach.
\section{Methodology}

\begin{figure}[t] 
	\centering
	\includegraphics[width=\textwidth]{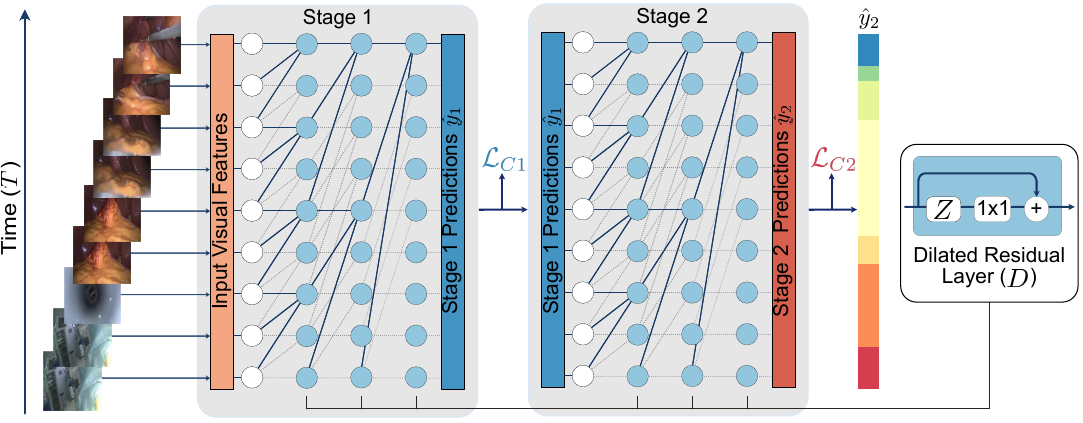}
	\caption{Overview of the proposed TeCNO multi-stage hierarchical refinement model. The extracted frame features are forwarded to Stage 1 of our TCN, which consists of 1D dilated convolutional and dilated residual layers $D$. Cross-Entropy Loss is calculated after each stage and aggregated for the joint training of the model.} 
	\label{fig1}
\end{figure}

TeCNO constitutes a surgical workflow recognition pipeline consisting of the following steps: 1) We employ a ResNet50 as a visual feature extractor. 2) We refine the extracted features with a 2-stage causal TCN model that forms a high-level reasoning of the current frame by analyzing the preceding ones. The refinement 2-stage TCN model is depicted in Fig.~\ref{fig1}.

 \subsection{Feature Extraction Backbone} 
 A ResNet50~\cite{He} is trained frame-wise without temporal context as a feature extractor from the video frames either on a single task for phase recognition or as a multi-task network when a dataset provides additional label information, for instance tool presence per frame. In the multi-task scenario for concurrent phase recognition and tool identification, our model concludes with two separate linear layers, whose losses are combined to train the model jointly. Since phase recognition is an imbalanced multi-class problem we utilize softmax activations and weighted cross entropy loss for this task. The class weights are calculated with median frequency balancing~\cite{mfb}. For tool identification, multiple tools can be present at every frame, constituting a multi-label problem, which is trained with a binary-cross entropy loss after a sigmoid activation.
 
 We adopt a two-stage approach so that our temporal refinement pipeline is independent of the feature extractor and the available ground truth provided in the dataset. As we will discuss in Section~\ref{results}, TCNs are able to refine the predictions of various features extractors regardless of their architecture and label information.
 
\subsection{Temporal Convolutional Networks} \label{Temporal Convolutional Networks}
For the temporal phase prediction task, we propose TeCNO, a multi-stage temporal convolutional network that is visualized in Fig.~\ref{fig1}.
Given an input video consisting of $ x_{1:t},\ 
t \in [1,T]$ frames, 
where $T$ is the total number of frames, the goal is to predict $y_{1:t}$ where $y_t$ is the 
class label for the current time step $t$.
Our temporal model follows the design of MS-TCN and contains neither pooling layers, that would decrease the temporal resolution, nor fully connected layers, which would increase the number of parameters and require a fixed input dimension. Instead, our model is constructed solely with temporal convolutional layers.

The first layer of Stage 1 is a 1x1 convolutional layer that matches the input feature dimension to the chosen feature length forwarded to the next layer within the TCN. Afterwards, dilated residual ($D$) layers perform dilated convolutions as described in Eq.~\ref{eqDilatedConvolution} and Eq.~\ref{eqDilatedResidualLayer}. The major component of each $D$ layer is the dilated convolutional layer ($Z$). 

\begin{align} \label{eqDilatedConvolution}
    Z_l &= ReLU(W_{1,l} *  D_{l-1} +b_{1,l}) \\
 \label{eqDilatedResidualLayer}
    D_l &= D_{l-1} + W_{2,l}*Z_l +b_{2,l}
\end{align}
$D_l$ is the output of $D$ (Eq. \ref{eqDilatedResidualLayer}), while $Z_l$ is the result of the dilated convolution of kernel $W_{1,l}$ with the output of the previous layer $D_{l-1}$ activated by a $ReLU$ (Eq. \ref{eqDilatedConvolution}). $W_{2,l}$ is the kernel for the 1x1 convolutional layer, $*$ denotes a convolutional operator and $b_{1,l}, b_{2,l}$ are bias vectors.

\begin{sloppypar}
Instead of the acausal convolutions in MS-TCN~\cite{Farha} with predictions
$\hat{y_t}(x_{t-n},...,x_{t+n})$ which depend on both $n$ past and $n$ future frames, we use causal convolutions within our $D$ layer. 
Our causal convolutions can be easily described as 1D convolutions with kernel size 3 with a dilation factor. The term causal refers to the fact that the output of each convolution is shifted and the prediction $\hat{y}$ for time step $t$ does not rely on any $n$ future frames but only relies on the current and previous frames i.e. $\hat{y_t}(x_{t-n},...,x_{t})$. This allows for intra-operative online deployment of TeCNO.
\end{sloppypar}

Increasing the dilation factor of the causal convolutions by 2 within the $D$ layer for each consecutive layer we effectively increase the temporal receptive field $RF$ of the network without a pooling operation (Eq.\ref{eqRF}). We visualize the progression of the receptive field of the causal convolutions in Fig.~\ref{fig1}.
A single $D$ layer with a dilation factor of 1 and a kernel size of 3 can process three time steps at a time. Stacking 3 consecutive $D$ layers within a stage, as seen in Fig.~\ref{fig1}, increases the temporal resolution of the kernels to 8 time steps. The size of the temporal receptive field depends on the number of $D$ layers $l \in [1,N]$ and is given by:
\begin{equation}
    RF(l) = (2)^{l+1}-1
    \label{eqRF}
\end{equation}

This results in a exponential increase of the receptive field, which significantly reduces the computational cost in comparison to models that achieve higher receptive field by increasing the kernel size or the amount of total layers~\cite{VanDenOord}.
 
\subsubsection{Multi-Stage TCN}
The main idea of the multi-stage approach is to refine the output of the first stage $S_1$ by adding $M$ additional stages to the network $S_{1...M}$~\cite{Newell}. The extracted visual feature vectors for each frame of a surgical video $x_{1:T}$ are the input of $S_1$, as explained above. The output of $S_1$ is directly fed into the second stage $S_2$. As seen in Fig.~\ref{fig1}, the outputs of $S_1$ and $S_2$ have independent loss functions and the reported predictions are calculated after $S_2$, where the final refinement is achieved. 

After each stage $S_{1...M}$ we use a weighted cross-entropy loss to train our model, as described in Eq.~\ref{multi_stage_ce}. Here, $y_t$ is the ground truth phase label and $\hat{y}_{mt}$ is the output prediction of each stage $m \in [1,M]$. The class weights $w_c$ are calculated using median frequency balancing~\cite{mfb} to mitigate the imbalance between phases. Our TeCNO model is trained utilizing exclusively phase recognition labels without requiring any additional tool information.
\begin{equation}
     \mathcal{L}_{C} = \dfrac{1}{M} \sum_{m}^{M}\mathcal{L}_{Cm} = - \frac{1}{M}\frac{1}{T} \sum_{m}^{M}\sum_{t}^{T}w_c y_{mt} \cdot log(\hat{y}_{mt})
     \label{multi_stage_ce}
\end{equation}
\section{Experimental Setup}
\noindent\textbf{Datasets}
We evaluated our method on two challenging surgical workflow intra-operative video datasets of laparoscopic cholecystectomy procedures for the resection of the gallbladder. The publicly available  Cholec80~\cite{Twinanda2016} includes 80 videos with resolutions 1920$\times$1080 or 854$\times$480 pixels recorded at 25 frames-per-second (fps). Each frame is manually assigned to one of seven classes corresponding to each surgical phase. Additionally, seven different tool annotation labels sampled at 1fps are provided. The dataset was subsampled to 5fps, amounting to $\sim$92000 frames. We followed the split of~\cite{Twinanda2017},~\cite{Jin2020} separating the dataset to 40 videos for training, 8 for validation, and 32 for testing.

Cholec51 is an in-house dataset of 51 laparoscopic cholecystectomy videos with resolution 1920$\times$1080 pixels and sampling rate of 1fps. Cholec51 includes seven surgical phases that slightly differ from Cholec80 and have been annotated by expert physicians. There is no additional tool information provided. 25 videos were utilized for training, 8 for validation and 18 for test. Our experiments for both datasets were repeated 5 times with random initialization to ensure reproducability of the results.

\subsubsection{Model Training}
TeCNO was trained for the task of surgical phase recognition using the Adam optimizer with an initial learning rate of 5e-4 for 25 epochs. We report the test results extracted by the model that performed best on the validation set. The batch size is identical to the length of each video.
Our method was implemented in PyTorch and our models were trained on an NVIDIA Titan V 12GB GPU using Polyaxon\footnote{https://polyaxon.com/}. The source code for TeCNO will become publicly available upon acceptance of the manuscript.

\subsubsection{Evaluation Metrics}
To comprehensively measure the results of the phase prediction we deploy three different evaluation metrics suitable for surgical phase recognition~\cite{Padoy2012}, namely Accuracy (Acc), Precision (Prec) and Recall (Rec). Accuracy quantitatively evaluates the amount of correctly classified phases in the whole video, while Precision, or positive predictive value, and Recall, or true positive rate, evaluate the results for each individual phase~\cite{Twinanda2017a}.

\subsubsection{Ablative Testing}
To identify a suitable feature extractor for our MS-TCN model we performed experiments with two different CNN architectures, namely AlexNet~\cite{Krizhevsky2012} and ResNet50~\cite{He}. Additionally we performed experiments with different number of TCN stages to identify which architecture is best able to capture the long temporal associations in our surgical videos.

\subsubsection{Baseline Comparison}
TeCNO was extensively evaluated against surgical phase recognition networks, namely, PhaseLSTM~\cite{Twinanda2017},  EndoLSTM~\cite{Twinanda2017} and MTRC-Net~\cite{Jin2020}, which employ LSTMs to encompass the temporal information in their models. We selected LSTMs over HMMs, since their superiority has been extensively showcased in the literature~\cite{Yengera2018}. Moreover, MTRCNet is trained in an end-to-end fashion, while the remaining LSTM approaches and TeCNO focus on temporally refining already extracted features. Since Cholec51 does not include tool labels, EndoLSTM and MTRCNet are not applicable due to their multi-task requirement. All feature extractors for Cholec80 were trained for a combination of phase and tool identification, except for the feature extractor of PhaseLSTM~\cite{Twinanda2016}, which requires only phase labels. The CNNs we used to extract the features for Cholec51 were only trained on phase recognition since no tool annotations were available.

\begin{table}[t]\centering
\caption{Ablative testing results for different feature extraction CNNs and increasing number of stages for Cholec80. Average metrics over multiple runs are reported (\%) along with their respective standard deviation.}
\resizebox{\textwidth}{!}{
\begin{tabular}{ @{} c *{6}{S[table-format=2.2,table-figures-uncertainty=4, detect-weight]} @{}}
\toprule
 & \multicolumn{3}{c}{\textbf{AlexNet}} & \multicolumn{3}{c@{}}{\textbf{ResNet50}} \\ 
\cmidrule(lr){2-4} \cmidrule(l){5-7}
& {Acc} & {Prec} & {Rec} & {Acc} & {Prec} & {Rec}  \\
\midrule
\textbf{No TCN}  & 74.40 \pm 4.30   & 63.06 \pm 0.32 & 70.75 \pm 0.05 & 82.22 \pm 0.60 & 70.65 \pm 0.08 & 75.88 \pm 1.35\\
\textbf{Stage I}  & 84.04 \pm 0.98   & 79.82 \pm 0.31 & 79.03 \pm 0.99 & 88.35 \pm 0.30 & \bfseries 82.44 \pm 0.46 & 84.71 \pm 0.71\\
\textbf{Stage II}   & 85.31 \pm 1.02   & 81.54 \pm 0.49 & 79.92 \pm 1.16 & \bfseries 88.56 \pm 0.27 & 81.64 \pm 0.41 & \bfseries 85.24 \pm 1.06 \\
\textbf{Stage III}   & 84.41 \pm 0.85   & 77.68 \pm 0.90 & 79.64 \pm 1.6 & 86.49 \pm 1.66 & 78.87 \pm 1.52 & 83.69 \pm 1.03 \\
\bottomrule
\end{tabular}
}
\label{tab:ablative_testing}
\end{table}
\section{Results} \label{results}

\subsubsection{Effect of Feature Extractor Architecture}
As can be seen in Table~\ref{tab:ablative_testing}, ResNet50 outperforms AlexNet across the board with improvements ranging from 2\% to 8\% in accuracy. Regarding precision and recall, the margin increases even further. For all stages ResNet50 achieves improvement over AlexNet of up to 7\% in precision and 6\% in recall. This increase can be attributed to the improved training dynamics and architecture of ResNet50~\cite{He}. Thus, the feature extractor selected for TeCNO is ResNet50.
\begin{table}[t]\centering
\caption{Baseline Comparison for Cholec80 and Cholec51 Datasets. EndoLSTM and MTRCNet require tool labels, therefore cannot be applied for Cholec51. The average metrics over multiple runs are reported (\%) along with their respective standard deviation.}
\resizebox{\textwidth}{!}{
\begin{tabular}{ @{} c *{6}{S[table-format=2.2,table-figures-uncertainty=4, detect-weight]} @{}}
\toprule
 & \multicolumn{3}{c}{\textbf{Cholec80}} & \multicolumn{3}{c@{}}{\textbf{Cholec51}} \\ 
\cmidrule(lr){2-4} \cmidrule(l){5-7}
& {Acc} & {Prec} & {Rec} & {Acc} & {Prec} & {Rec}  \\
\midrule
PhaseLSTM~\cite{Twinanda2017}  & 79.68\pm 0.07   & 72.85\pm 0.10 & 73.45\pm 0.12 & 81.94\pm 0.20 & 68.84\pm 0.11 & 68.05\pm 0.79\\
EndoLSTM~\cite{Twinanda2017a} & 80.85\pm 0.17   & 76.81 \pm 2.62 & 72.07\pm 0.64 & \textemdash & \textemdash & \textemdash \\
MTRCNet~\cite{Jin2020}   & 82.76\pm 0.01   & 76.08\pm 0.01 & 78.02\pm 0.13 & \textemdash & \textemdash & \textemdash \\
ResNetLSTM~\cite{Jin2018}   & 86.58\pm 1.01   & 80.53\pm 1.59 & 79.94\pm 1.79 & 86.15\pm 0.60 & 70.45\pm 2.85 & 67.42\pm 1.43 \\
TeCNO   & \bfseries 88.56\pm 0.27   & \bfseries 81.64\pm 0.41 & \bfseries 85.24\pm 1.06 & \bfseries 87.34\pm 0.66 & \bfseries 75.87\pm 0.58 & \bfseries 77.17\pm 0.73 \\
\bottomrule
\end{tabular}
}
\label{tab:baselines}
\end{table}

\begin{figure}[t]
	\centering
	\includegraphics[width=\textwidth]{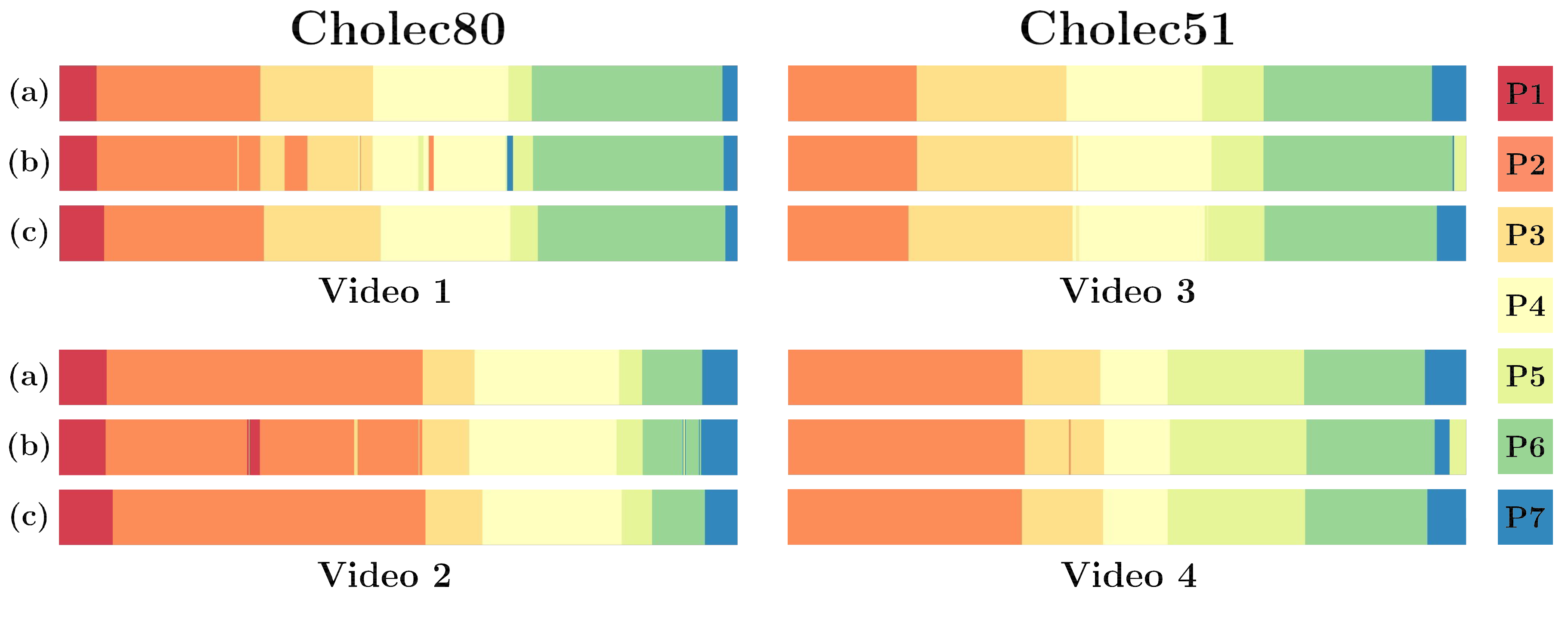}
	\caption{Qualitative Results regarding quality of phase recognition for Cholec80 and Cholec51. (a) Ground Truth (b) ResNetLSTM Predictions (c) TeCNO Predictions. P1 to P7 indicate the phase label.
	} 
	\label{qualitative_results}
\end{figure}

\subsubsection{Effect of TCN and Number of Stages}
Table~\ref{tab:ablative_testing} also highlights the substantial improvement in the performance achieved by the TCN refinement stages. Both AlexNet and ResNet50 obtain higher accuracy by 10\% and 6\% respectively with the addition of just 1 TCN Stage. Those results signify not only the need for temporal refinement for surgical phase recognition but also the ability of TCNs to improve the performance of any CNN employed as feature extractor, regardless of its previous capacity. We can also observe that the second stage of refinement improves the prediction of both architectures across our metrics. However, Stage 2 outperforms Stage 3  by 1\% in accuracy for AlexNet and 2\% for ResNet50. This could indicate that 3 stages of refinement lead to overfitting on the training set for our limited amount of data.

\subsubsection{Comparative Methods}
In Table 2 we present the comparison of TeCNO with different surgical phase recognition approaches that utilize LSTMs to encompass the temporal information in their predictions. PhaseLSTM~\cite{Twinanda} and EndoLSTM~\cite{Twinanda} are substantially outperformed by ResNetLSTM and TeCNO by 6\% and 8\% in terms of accuracy for both datasets respectively.
This can be justified by the fact that they employ AlexNet for feature extraction, which as we showed above has limited capacity.
Even though MTRCNet is trained in an end-to-end fashion, it is also outperformed by 4\% by ResNetLSTM and 6\% by TeCNO, which are trained in a two-step process. Comparing our proposed approach with ResNetLSTM we  notice an improvement of 1-2\% in accuracy. However, the precision and recall values of both datasets are substantially higher by 6\%-10\%. The higher temporal resolution and large receptive field of our proposed model allow for increased performance even for under-represented phases. 

\subsubsection{Phase Recognition Consistency}
In Fig.~\ref{qualitative_results} we visualize the predictions for four laparoscopic videos, two for each dataset. The results clearly highlight the ability of TeCNO to obtain consistent and smooth predictions not only within one phase, but also for the often ambiguous phase transitions. Compared against ResNetLSTM, TeCNO can perform accurate phase recognition, even for the phases with shorter duration, such as P5 and P7. Finally, TeCNO showcases robustness, since Video 3 and 4 are both missing P1. However, the performance of our model does not deteriorate.
\section{Conclusion}
In this paper we proposed TeCNO, a multi-stage Temporal Convolutional Neural Network, which was successfully deployed on the task of surgical phase recognition. Its full temporal resolution and large receptive field allowed for increased performance against a variety of LSTM-based approaches across two datasets. Online and fast inference on whole video-sequences was additionally achieved due to causal, dilated convolutions. The multi-stage refinement process increased the prediction consistency, not only within phases, but also in the ambiguous inter-phase transitions. Future work includes evaluation of our method on a larger number of videos from a variety of laparoscopic procedures.

\section*{Acknowledgements}
\label{sec:acknowledgements}
Our research is funded by the DFG research unit 1321 PLAFOKON  in collaboration with the Minimal-invasive Interdisciplinary Intervention Group (MITI). We would also like to thank NVIDIA for the GPU donation.

\bibliographystyle{ieeetr}
\bibliography{bibfile}

\end{document}